\documentclass{article}

\usepackage{arxiv}
\usepackage[utf8]{inputenc} 
\usepackage[T1]{fontenc}    
\usepackage{url}            
\usepackage{booktabs}       
\usepackage{amsfonts}       
\usepackage{nicefrac}       
\usepackage{microtype}      
\usepackage{lipsum}		
\usepackage{graphicx}
\usepackage{titlesec}
\usepackage{achemso} 
\setkeys{acs}{usetitle = true}
\usepackage{multirow}
\usepackage{textcomp}
\usepackage{float}          
\usepackage[version=3]{mhchem}
\usepackage{doi}
\usepackage{chngcntr}
\counterwithin{figure}{section}
\usepackage{ragged2e}
\usepackage{anyfontsize}
\usepackage{graphicx}
\usepackage{amsmath}
\usepackage{gensymb}
\usepackage{newtxtext} 
\usepackage{newtxmath} 
\usepackage{multicol}
\usepackage{hyperref}       
\usepackage{caption}        
\usepackage{xcolor}
\usepackage{siunitx}

\renewcommand{\footnotesize}{\fontsize{7.68pt}{9.22pt}\selectfont}
\renewcommand{\scriptsize}{\fontsize{6.72pt}{8.06pt}\selectfont}
\renewcommand{\tiny}{\fontsize{6.24pt}{7.49pt}\selectfont}
\renewcommand{\small}{\fontsize{8.64pt}{10.37pt}\selectfont}
\renewcommand{\normalsize}{\fontsize{10.24pt}{12.29pt}\selectfont}
\renewcommand{\large}{\fontsize{13.44pt}{16.19pt}\selectfont}
\renewcommand{\Large}{\fontsize{16.13pt}{19.43pt}\selectfont}
\renewcommand{\LARGE}{\fontsize{19.35pt}{23.22pt}\selectfont}
\renewcommand{\huge}{\fontsize{22.74pt}{27.29pt}\selectfont}
\renewcommand{\Huge}{\fontsize{27.29pt}{32.81pt}\selectfont}

\title{Propensity of water self-ions at air(oil)-water interfaces revealed by deep potential molecular dynamics with enhanced sampling}

\author{
    \begin{center}
        \textbf{Pengchao Zhang}\textsuperscript{1}, and \textbf{Xuefei Xu}\textsuperscript{1,*}.\\[0.5em]
        \textsuperscript{1}Center for Combustion Energy, Department of Energy and Power Engineering, and Key Laboratory for Thermal Science and Power Engineering of Ministry of Education, Tsinghua University, Beijing 100084, China\\[0.5em]    
        \textsuperscript{*}Corresponding author e-mail: xuxuefei@tsinghua.edu.cn
    \end{center}
}

\date{}

\renewcommand{\headeright}{}
\renewcommand{\undertitle}{}
\renewcommand{\shorttitle}{Zhang et. al., Propensity of water self-ions at air(oil)-water interfaces}

\titlespacing{\section}{0pt}{\parskip}{-\parskip}
\titlespacing{\subsection}{0pt}{\parskip}{-\parskip}

\DeclareUnicodeCharacter{0308}{\"{o}}
\DeclareUnicodeCharacter{0301}{\'}

\begin{document}
\maketitle

\begin{abstract}
The preference of water self-ions (hydronium and hydroxide) towards air/oil-water interfaces is one of the hottest topics in water research due to its importance for understanding properties, phenomena, and reactions of interfaces. In this work, we performed enhanced-sampling molecular dynamics simulations based on state-of-the-art neural network potentials with approximate M06-2X accuracy to investigate the propensity of hydronium and hydroxide ions at air/oil(decane)-water interfaces, which can simultaneously describe well the water autoionization process forming these ions, recombination of ions, and ionic distribution along the normal distance to the interface by employing a set of appropriate Voronoi collective variables. 
A stable ionic double-layer distribution is observed near the air-water interface, while the distribution is different at oil-water interfaces, where hydronium tends to be repelled from the interface into the bulk water, whereas hydroxide, with an interfacial stabilization free energy of --0.6 kcal/mol, is enriched in the interfacial layer. Through simulations of oil droplets in water, we further reveal that the interfacial propensity of hydroxide ions is caused by the positive charge distribution of the oil-water interface contributed by hydrogens of the dangling OH bonds of interfacial water layer and the outmost layer decane molecules laying flat on the droplet. The present results may aid in understanding the acid-base nature of water interfaces with wide applications.
\end{abstract}

\begin{multicols}{2}
\renewcommand{\thefigure}{\arabic{figure}}
\justifying

\section*{1 Introduction}
Water interfaces play a crucial role in both natural sciences and engineering applications due to providing fundamental heterogeneous environments.\cite{ruiz2020molecular,wei2020accelerated,björneholm2016water} The preferential accumulation of water self-ions at these interfaces, namely hydroxide (\ce{OH-}) and hydronium (\ce{H3O+}), can impact both interfacial properties\cite{chamberlayne2020simple,martins2023electrostatics} and chemical reaction processes,\cite{lee2019spontaneous,zhong2022determining,martins2023electrostatics} and has been widely investigated. Nevertheless, the propensity of self-ions enriching at air-water and oil-water interfaces continues to be an open question due to contradictory findings from these previous studies.\cite{vacha2007autoionization,buch2007water,beattie2009surface,mishra2012bronsted,saykally2013two}

For example, the macroscopic experiments measured the negative zeta potential of air bubbles\cite{quincke1861ueber,graciaa1995zeta,creux2009strong,creux2007specific} and oil droplets\cite{creux2009strong,marinova1996charging,beattie2004pristine} in water and inferred the enrichment of \ce{OH-} at the air(oil)-water interfaces considering that \ce{OH-} is the only anionic source in pure water. The second harmonic generation\cite{fang2015evidence} and sum-frequency vibrational spectroscopy\cite{Tian_ProcNatlAcadSciUSA_2009_v106_p15148,yang_stabilization_2020} studies also provided microscopic evidence for the adsorption of \ce{OH-} at the oil-water interface. However, spectroscopic measurements of air-water interfaces observed the interfacial enhancement of \ce{H3O+}.\cite{vacha2007autoionization,petersen2005evidence,das2019nature,chiang2020affinity} Similarly, microscopic simulations also obtained conflicting conclusions on the interfacial excess of self-ions. Most microscopic studies found that \ce{H3O+} ions tend to accumulate at the top layer of air-water interfaces,\cite{petersen2004hydrated,buch2007water,lee2009ab,hub2014thermodynamics,tse2015propensity,mamatkulov2017orientation,li2020coordinate} while \ce{OH-} ions are repelled into the water bulk\cite{hub2014thermodynamics,tse2015propensity} or show no preference.\cite{de2023acidity} A few studies observed that \ce{OH-} ions, rather than \ce{H3O+}, slightly accumulate at the air-water interfaces.\cite{baer2014toward,mundy2009hydroxide,bai2016surface}

Our recent deep potential molecular dynamics (DPMD) simulations on ionic distribution provide a possible explanation for the inconsistency of these conclusions at the air-water interface\cite{zhang2023double} and nitrogen-water interface.\cite{Zhang_JAmChemSoc_2024_v146_p19537} We found an interfacial double-layer distribution of water self-ions.\cite{zhang2023double}
In this distribution, \ce{H3O+} predominantly occupies the top layer of the interface, while \ce{OH-} is enriched in the deeper interfacial layer below that of \ce{H3O+}. Compared to \ce{H3O+},  \ce{OH-} has a more negative free energy of interfacial stabilization and accordingly a higher interfacial concentration, leading to negatively charged interfaces and negative zeta potential. 
These results are consistent with the macroscopic detection and do not contradict microscopic observation, as the previous conflicting results may be caused by measuring at different or ambiguous interfacial depths. And this finding has been further evidenced by a high-level heterodyne-detected vibrational sum frequency generation (vSFG) study.\cite{Yair2024Surface} However, it is important to note that our previous work performed separate simulations for \ce{OH-} and \ce{H3O+}, without considering their origins and recombination at the interface. Therefore, in this work, we performed further investigation with the corresponding consideration to validate the observation of the double-layer distribution.

Technically, it is pretty demanding to simulate the water autoionization reaction of forming \ce{OH-} and \ce{H3O+}, because it is an exceedingly rare event. Dynamics studies on the process in bulk water have been conducted for decades.\cite{eigen1955untersuchungen,eigen1958self,aaqvist1993simulation,trout1998dissociation,geissler2001autoionization,hassanali2011recombination,moqadam2018local} 
Recently, there was significant progress in accurate simulations of the water autoionization reaction and \ce{pKw},\cite{di2023water,liu2023mechanistic,calegari2023probing,dasgupta2024nuclear} benefiting primarily from the developments in neural network potentials and enhanced sampling methods.
Here, we utilized the on-the-fly probability enhanced sampling (OPES) method\cite{invernizzi2020rethinking,invernizzi2022exploration} with a set of Voronoi collective variables (CVs)\cite{grifoni2019microscopic,grifoni2020tautomeric,zhang2023intramolecular} to sample this rare event efficiently, and we employed neural network potential models with approximate M06-2X\cite{zhao2008m06} accuracy to drive MD simulations. Armed with these state-of-the-art methods, we obtained the free energy of water autoionization in quantitative agreement with experimental observations. Most significantly, the present deep-going investigation confirms the phenomenon of the ionic double-layer distribution at the air-water interface with the full consideration of water autoionization and its reverse recombination processes. 
In addition, we extended our study to oil-water (decane-water) interfaces, and observed a totally distinct distribution of water self-ions, where \(\ce{OH-}\) ions preferentially accumulate near the decane-water interface, while \(\ce{H3O+}\) ions are repelled into the bulk.

\section*{2 Results and discussion}

\subsection*{2.1 Water autoionization}
\begin{figure*}
  \centering
  \includegraphics[width=\textwidth]{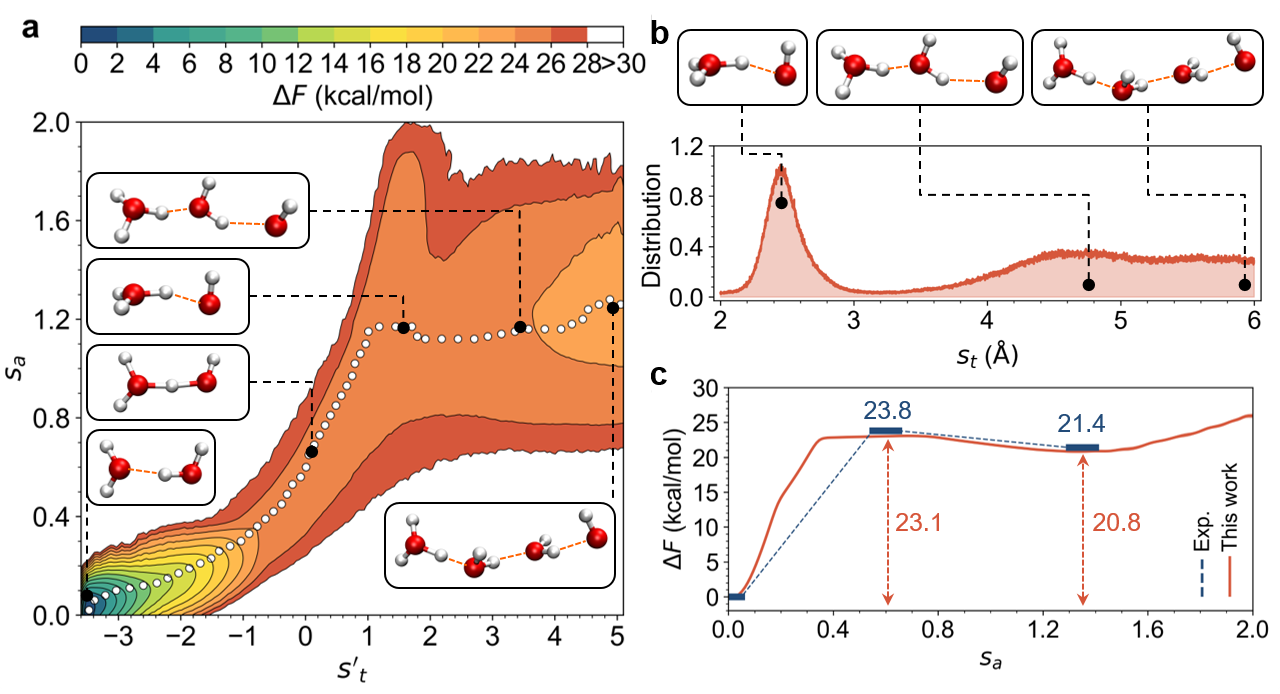}
  \caption{\textbf{Free energy surface of water autoionization.} \textbf{a}. Two-dimensional FES as a function of the logarithmic variant of ionic distance  $\mathbf{s}'_{t}$  and the ionic number $\mathbf{s}_a$, where the minimum free energy path of autoionization is represented by the white dot line. Key structures along the potential optimal pathway of water autoionization are provided, in which surrounding water molecules are hidden. Color code: hydrogen, white; oxygen, red. \textbf{b}. Distribution of self-ion distance $\mathbf{s}_t$ in product region and corresponding structures of ion pairs. \textbf{c}. One-dimensional projection of the FES along the ionic number coordinate $\mathbf{s}_a$. Experimental values are also provided for comparison.}
  \label{fig_ionize_geo}
\end{figure*}

Before discussing ionic preferences at interfaces, it is important to validate the feasibility of our methodologies for accurately describing the water autoionization and its reverse process. We first need an efficient potential with an accuracy approaching hybrid meta-GGA functional M06-2X.\cite{zhao2008m06} We chose to learn the energies and forces of the M06-2X functional because of its reported accuracy in modeling liquid water\cite{villard2023structure} and van der Waals interactions.\cite{giroday2014thermodynamic} To expedite this step, we initially trained a deep Kohn-Sham (DeePKS) model\cite{chen2020deepks,chen2023deepkscpc,li2022deepksabacus} with the precision of M06-2X, which saves approximately nine times more time than the regular M06-2X calculations. Then, we trained a deep potential (DP) model using the DeePMD-kit package\cite{wang2018deepmdkit,zeng2023deepmd} to closely match the accuracy of the DeePKS model, i.e., the trained DP model would have an approximate M06-2X accuracy. Finally, molecular dynamics simulations were run by using the DP model,  which offers a time-saving of more than five orders of magnitude when compared to the conventional \textit{ab initio} MD (AIMD) simulations directly using the M06-2X. 

In the simulation, to efficiently sample the rare event of water autoionization, we employed the OPES method. In this way, appropriate CVs must be defined. To depict both the production of \ce{OH-} and \ce{H3O+} by the water autoionization and the diffusion of these self-ions through the Grotthuss mechanism,\cite{von1805memoire,marx2006proton,zhang2023double} we tessellated the space into Voronoi polyhedra centered on all oxygen atoms and summed up the atomic charges within each polyhedron,\cite{grifoni2019microscopic,grifoni2020tautomeric,zhang2023intramolecular} that can enable us to automatically identify charge defects of \ce{OH-} and \ce{H3O+} without a priori knowledge, and accordingly designed CVs, which are self-ion number $\mathbf{s}_a$ 
\begin{equation}
    \mathbf{s}_a = \sum_{i=1}^{\mathrm{Num_{O}}} \delta_{i}^2,
\label{eq:s_a}
\end{equation}
self-ion distance $\mathbf{s}_t$
\begin{equation}
    \mathbf{s}_t = -\sum_{i=1}^{\mathrm{Num_{O}}} \sum_{j>i}^{\mathrm{Num_{O}}} r_{i, j} \delta_{i} \delta_{j} ,
\label{eq:s_t}
\end{equation}
and its variant $\mathbf{s}'_t$
\begin{equation}
    \mathbf{s}'_{t} = \begin{cases}
\log(\mathbf{s}_t + \epsilon), & \ 0 \leq \mathbf{s}_t < 1, \\
\mathbf{s}_t - 1 + \log(1+\epsilon), & \ \mathbf{s}_t \geq 1,
\end{cases}
\label{eq:s_logt}
\end{equation}
to distinguish the water autoionization process. 
The charge defect of the \textit{i}th Voronoi center (O atom) is denoted by $\delta_{i}$, with values 0, 1, and --1 for \ce{H2O}, \ce{H3O+}, and \ce{OH-} respectively. Distance between two Voronoi centers is represented by $r_{i,j}$. Consequently, the CV $\mathbf{s}_a$ signifies the number of \ce{H3O+} and \ce{OH-} ions resulting from water autoionization, while the CV $\mathbf{s}_t$ indicates the distance between two ions. However, the CV $\mathbf{s}_t$ struggles to differentiate between pure water and transition states, so we introduced its logarithmic variant CV $\mathbf{s}'_t$ to address this issue. In the definition of  $\mathbf{s}'_t$ (eq \ref{eq:s_logt}), the $\epsilon $ is a regularization parameter and is set as 0.03. More details of CVs and other simulation settings are provided in the Supporting Information (SI).

Based on the DPMD simulations, we calculated the free energy surface (FES) as a function of the two defined CVs ($\mathbf{s}_a$ and $\mathbf{s}'_t$). As shown in Figure \ref{fig_ionize_geo}a, a deep well is located at $\mathbf{s}_a = 0$ and $\mathbf{s}'_t = -3.6$ ($\mathbf{s}_t = 0$ Å) which presents the neutral water molecules bonded with the hydrogen bonds, namely, the reactant of water autoionization reaction. We also observed a shallow well at $\mathbf{s}_a = 1.3$ and $\mathbf{s}'_t = 4.8$ corresponding to the autoionization products (\ce{OH-} and \ce{H3O+}) separated by a water wire with a length of $\sim$6 Å that indicates the metastability of self-ions pair.\cite{natzle1985recombination,liu2023mechanistic} We further identified the optimal path connecting the two wells and marked it with the white dot line in Figure \ref{fig_ionize_geo}a, accompanied by the key structures along the path. It is revealed that the water autoionization occurs via the concerted transfer of multiple protons along the water wire, which follows the Grotthuss mechanism,\cite{hassanali2011recombination,moqadam2018local} involves sequentially a contact \ce{OH-} and \ce{H3O+} ion pair (corresponding to a self-ion distance $\mathbf{s}_t = 2.5$ Å in Figure \ref{fig_ionize_geo}b) and a one \ce{H2O}-bridged ion pair ($\mathbf{s}_t = 4.8$ Å in Figure \ref{fig_ionize_geo}b), and  eventually forms the autoionization products, \ce{OH-} and \ce{H3O+} ions separated by a water wire longer than  $\sim$6 Å. 

By projecting the FES along the self-ion number CV, $\mathbf{s}_a$, we got a one-dimensional free energy curve (Figure \ref{fig_ionize_geo}c) to clearly show the free energy change of the reaction process. From Figure \ref{fig_ionize_geo}c, the activation free energy ($\Delta F^{\ddag}$) and the reaction free energy ($\Delta F{\degree}$) of the water autoionization reaction are estimated as 23.1 and  20.8 kcal/mol, respectively, which are in good agreement with the experimental measurements ($\Delta F^{\ddag}$ = 23.8 kcal/mol,  $\Delta F{\degree}$ = 21.4 kcal/mol ).\cite{eigen1955untersuchungen,aaqvist1993simulation,trout1998dissociation}. The reverse process---recombination of self-ions, may occur by overcoming a low barrier of approximately 2.3 kcal/mol, which is close to the proton transfer barrier (2.4 kcal/mol) observed in NMR experiments.\cite{luz1964activation} These results demonstrate the accuracy of our simulations in characterizing the potential energy landscape of water autoionization.

\subsection*{2.2 Preference of water self-ions towards air(oil)-water slab interfaces}
\begin{figure*}
  \centering
  \includegraphics[width=\textwidth]{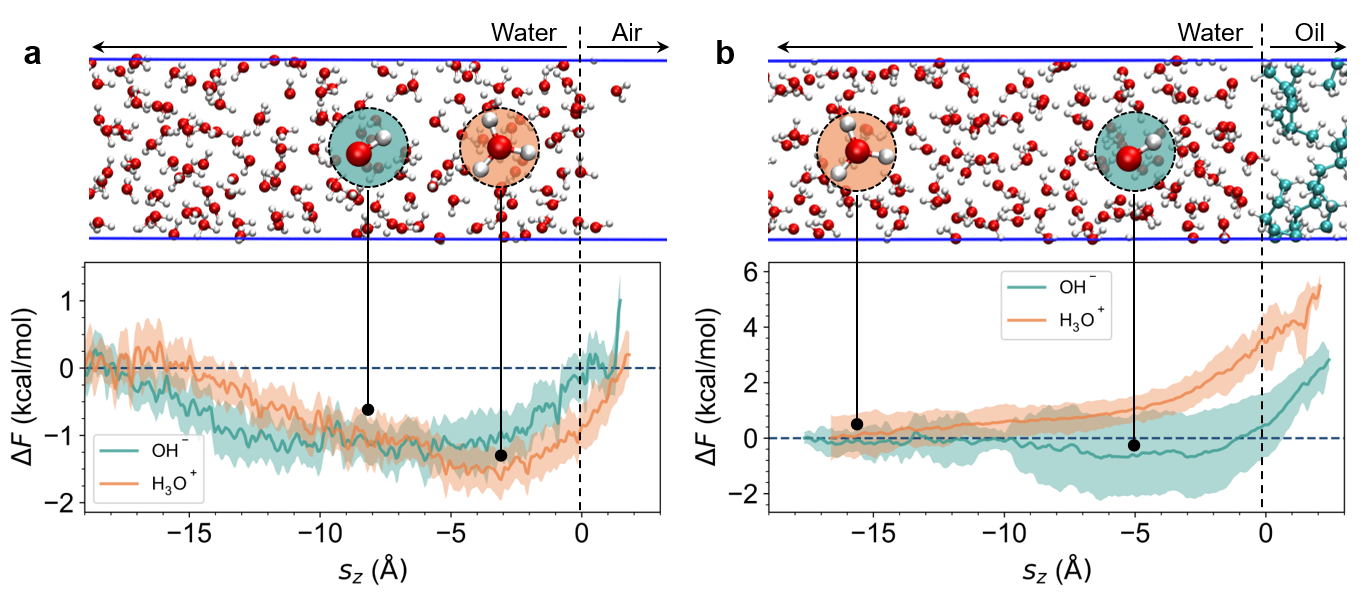}
  \caption{\textbf{Distribution of water self-ions near the interface.} One-dimensional free energy curves along the normal distance to \textbf{a}. the air-water slab interface and \textbf{b}. the oil-water slab interface are depicted below the corresponding schematic diagrams. The values of CV $\mathbf{s}_{z}$, representing the normal distances to the reference surface, are negative on the water side, zero at the surface, and positive on the air or oil side. In the schematic diagrams, particular emphasis is placed on highlighting the distinct preferential distribution of \ce{OH-} and \ce{H3O+} ions near the interface.  Color code: hydrogen, white; oxygen, red; carbon, green. }
  \label{fig_oil_air}
\end{figure*}

Having validated the accuracy of the methodologies, we shifted our focus to investigating the distribution preference of water self-ions near the air-water and oil-water (decane-water) interfaces. Note that we consider pure water without directly adding any ions in these simulations, and the self-ions are the products of water autoionization. To this end, in the simulations we introduced a new CV $\mathbf{s}_{z}$ 
\begin{equation}
    \mathbf{s}_{z} = \sum_{i=1}^{\mathrm{Num_{O}}} \delta z_i \delta_{i}^2 ,
\label{eq:s_z}
\end{equation}
where $\delta z$ indicates the relative coordinate along the z-axis. Thus, the CV $\mathbf{s}_{z}$ can describe depths of \ce{OH-} (for $\delta_{i} < 0$) and \ce{H3O+} (for $\delta_{i} > 0$) relative to the interfaces. 
The free energy curves of \ce{OH-} and \ce{H3O+} distribution as the function of $\mathbf{s}_{z}$ were calculated and plotted in Figure \ref{fig_oil_air}, where we took the position of the interface as the zero point of $\mathbf{s}_{z}$ for a better understanding of the relative locations to the interface. In this work, the interface of the air-water system is defined to be the surface where the water density is half of its bulk value, and the interface of the oil-water system is set as the  surface with the minimum density of the system considering the hydrophobicity of oil. Negative values and positive values of $\mathbf{s}_{z}$ denote the water phase and air (or oil) phase, respectively. 

Figure \ref{fig_oil_air}a supports the double-layer distribution of water self-ions near the air-water interface. We can clearly see two free energy wells located at different depths near the interface, which represent the preferential locations of \ce{OH-} and \ce{H3O+}, respectively. 
The free energy well of \ce{H3O+} spans the region (--15 {\AA} < $\mathbf{s}_{z}$ < 1 {\AA}) near the interface, with the minimum (--1.6 kcal/mol) located at $\mathbf{s}_{z}$ = --3 {\AA}; the free energy well of \ce{OH-} appears in a deeper layer spanning a wider region (--18 {\AA} < $\mathbf{s}_{z}$ < 0 {\AA}) below the interface, and the corresponding minimum has a free energy of --1.3 kcal/mol located at $\mathbf{s}_{z}$ = --8 {\AA}. (Note, the reported free energies here are the relative values to the bulk.) The energy well of \(\ce{H3O+}\) near the air-water interface is slightly deeper than that of \(\ce{OH-}\), with a difference of 0.3 kcal/mol at the air-water interface, that is opposite to the earlier work\cite{zhang2023double} which yielded a difference of --0.3 kcal/mol between the interfacial stabilization energies of \(\ce{H3O+}\) and \(\ce{OH-}\).  This discrepancy may arise from the improved sampling methods and potential accuracy in the present work. Nevertheless, the current simulation results reproduce the double-layer distribution of \(\ce{H3O+}\) and \(\ce{OH-}\) reported in the earlier work,\cite{zhang2023double} as well as their relative locations.

The distribution of water self-ions at the oil-water interface differs from that at the air-water interface, as shown in Figure \ref{fig_oil_air}b. The \(\ce{H3O+}\) ions dislike the oil-water interface, with the gradual free energy increases up to 3.4 kcal/mol from the bulk water ($\mathbf{s}_{z}$ = --16 {\AA}) to the interface ($\mathbf{s}_{z}$ = 0 {\AA}). In contrast, the free energy well of \(\ce{OH-}\) appears in the region (--9 {\AA} < $\mathbf{s}_{z}$ < --1 {\AA}) near the interface, with the corresponding minimum of --0.6 kcal/mol located at $\mathbf{s}_{z}$ = --5 {\AA}. This observation could account for the frequent detection of negative zeta potentials in cases involving oil droplets.\cite{creux2009strong,marinova1996charging,beattie2004pristine,noam2016protons} 

Combining the above simulation results of air-water and oil-water interfaces, we speculate that the experimental measurements of ionic preference at water interfaces using various experimental spectroscopic techniques\cite{du1994vibrational,inoue20152d,das2019nature,chiang2020affinity,yang_stabilization_2020} contradicted each other that might be caused by that they were probing different depths of the interfacial region or by the insufficient resolution. 
For instance, vSFG techniques, which provide information solely in the noncentrosymmetric region,\cite{ishiyama2017computational} may not effectively detect the distribution of ions in the deeper interfacial layer and specify the interfacial depth.\cite{tang2020molecular,sakunkaewkasem2023sum} 
\textit{Ab initio} MD simulations are usually too time-consuming to afford calculating on scales larger than nanometers, while the simulations using small periodic boxes may lead to a loss of information. Our previous DPMD simulations have addressed this issue by achieving \textit{ab initio} level accuracy with linear scaling relative to the number of atoms;\cite{zhang2023double,Zhang_JAmChemSoc_2024_v146_p19537} however, they still encounter challenges in fully sampling water self-ions in neutral systems. The present work employs enhanced sampling DPMD simulations to effectively tackle these challenges, enabling us to accurately identify the distinct distributions of water self-ions along the interfacial depth.

\subsection*{2.3 Distribution of water self-ions near the oil droplet in water}

\begin{figure*}
  \centering
  \includegraphics[width=\textwidth]{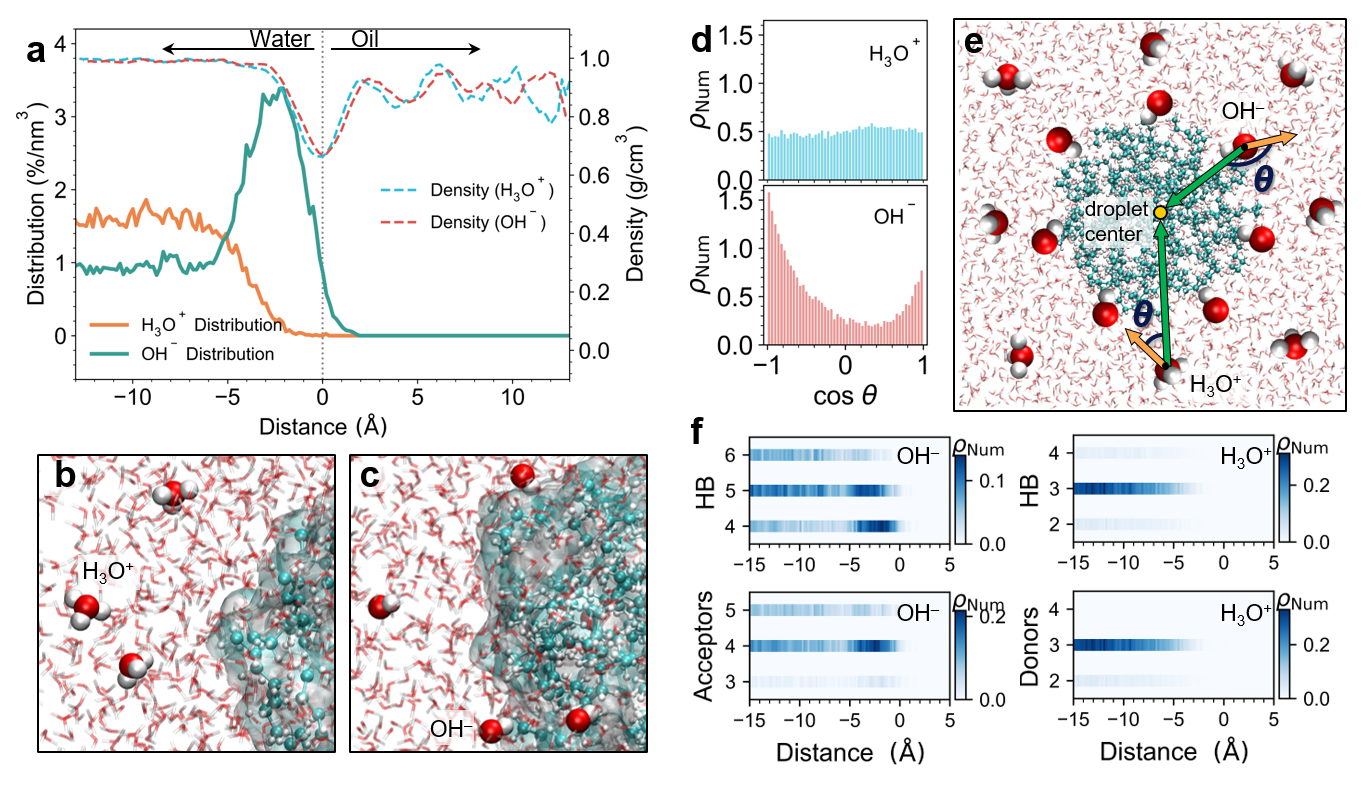}
  \caption{\textbf{Distribution and hydrogen bonding network of water self-ions near the oil-water interface.} \textbf{a}. The volume-normalized number distribution (solid lines) of  \ce{OH-} and \ce{H3O+} ions along the normal distance to the oil-water interface and the total density of systems (dashed lines) that contain \ce{OH-} or \ce{H3O+} ions. The spherical surface with the minimum density of system is defined as the oil-water interface. The normal distance is set as negative in the water region, zero at the oil-water interface, and positive in the oil droplet region. \textbf{b}. and \textbf{c}. Typical snapshots showing the repulsion of \ce{H3O+} and the accumulation of \ce{OH-} ions near the oil-water interface. \textbf{d.} Orientation distributions of \ce{OH-} (counted from --5 to 0 Å) and \ce{H3O+} (counted from --12 to --7 Å) in their preferred water layers. The histogram bars represent the cumulative number density, $\rho_\mathrm{Num}$, of self-ions within statistical intervals. The orientation of the ions is quantified by the cosine of the $\theta$ angle. \textbf{e.} Schematic representation of the $\theta$ angle between the normal vector (directing from water self-ions to the droplet center, indicated by green arrows) and the dipole vector (directing from the H center to O, illustrated by orange arrows) of the self-ions. \textbf{f.} The HB number per ion, as well as the number of HBs acting as acceptors or donors. The color bars indicate the cumulative number density \(\rho_\mathrm{Num}\) within statistical intervals.}
  \label{fig_distri_rdf_orien_hb}
\end{figure*}

To explain why water self-ions exhibit different preferences near the oil-water interface compared to the air-water interface, we constructed an oil (decane) droplet with a radius of \textasciitilde14 Å surrounded by over 5,000 water molecules to gather more information on the nanometer spatial scale. We conducted separate simulations for the oil droplet systems containing \(\ce{H3O+}\) and \(\ce{OH-}\) ions, since this approach can provide at least qualitatively consistent results with simulations using the OPES method and Voronoi CVs for ionic distribution at the air-water interface as validated in the above section and the latter is time-consuming when dealing with thousands of water molecules.

The spatial distribution analysis of \(\ce{H3O+}\) and \(\ce{OH-}\) ions in water surrounding the oil droplet is illustrated in Figure \ref{fig_distri_rdf_orien_hb}a. This analysis shows that \(\ce{OH-}\) ions preferentially accumulate near the oil-water interface, with a pronounced peak of the volume-normalized number distribution in the region from --5 to 0 Å relative to the interface.  Figure \ref{fig_distri_rdf_orien_hb}c shows a representative snapshot that illustrates the enrichment of \(\ce{OH-}\) ions near the oil-water interface.
In contrast, \(\ce{H3O+}\) ions exhibit a flat distribution from --12 to --7 Å, showing a weak propensity to be away from the oil-water interface (Figure \ref{fig_distri_rdf_orien_hb}b). As we have expected, these results are consistent with our free energy calculations based on the enhanced sampling DPMD simulation in an oil-water slab system in the section 2.2, which reveal that only \(\ce{OH-}\) ions have a preference towards the oil-water interface.

We further analyze the ionic orientation and hydrogen bonding network in the oil droplet system. For \(\ce{H3O+}\) ions preferentially distributed in the bulk water, their dipole vectors are distributed randomly, with \( \cos \theta \) ranging from --1.0 to 1.0, where the $\theta$  is the angle between the normal vector of interface and the ionic dipole vector (Figure \ref{fig_distri_rdf_orien_hb}d,e). 
The hydrogen bond (HB) number per \ce{H3O+} ion in water is primarily three, that is, the three hydrophilic protons of \ce{H3O+} act as donors, forming three HBs with neighboring water molecules (Figure \ref{fig_distri_rdf_orien_hb}f). 
Figure \ref{fig_distri_rdf_orien_hb}d shows that \(\ce{OH-}\) ions enriched in the oil-water interface are predominantly aligned anti-parallel to the interfacial normal vector, with \( \cos \theta \) around --1.0, indicating that interfacial \(\ce{OH-}\) ions are mainly dangling on the water surface with hydrogen towards the oil side. Thus, the hydrogen cannot act as an HB donor anymore, and the interfacial \(\ce{OH-}\) forms fewer hydrogen bonds (around 4, mainly by oxygen with more lone pair electrons as HB acceptors) as compared to those (around 5) distributed in the water bulk, as shown in Figures \ref{fig_distri_rdf_orien_hb}c,f).
Yang et al.\cite{yang_stabilization_2020} have ever observed the accumulation of \(\ce{OH-}\) ions at the hydrophobic hexane-water interface but they attributed the interfacial stability of \(\ce{OH-}\) to hexane-incorporated HB network, in which a hypercoordinated structure of  \(\ce{OH-}\) is formed with a larger average number of HBs  than \(\ce{OH-}\)  in the bulk water.  In contrast, we barely observe the weak hydrogen bonds formed between decane and \(\ce{OH-}\) ions in the present decane droplet system. This inconsistency may arise from the constructed model systems. Yang et al.\cite{yang_stabilization_2020} modeled a monolayer hexane gas-water system where the hexane molecules are adsorbed on the water surface by the Van der Waals interaction, while we modeled a decane droplet in water where the surface tension of droplet may affect the interfacial molecular interaction.

\begin{figure*}
  \centering
  \includegraphics[width=0.45\textwidth]{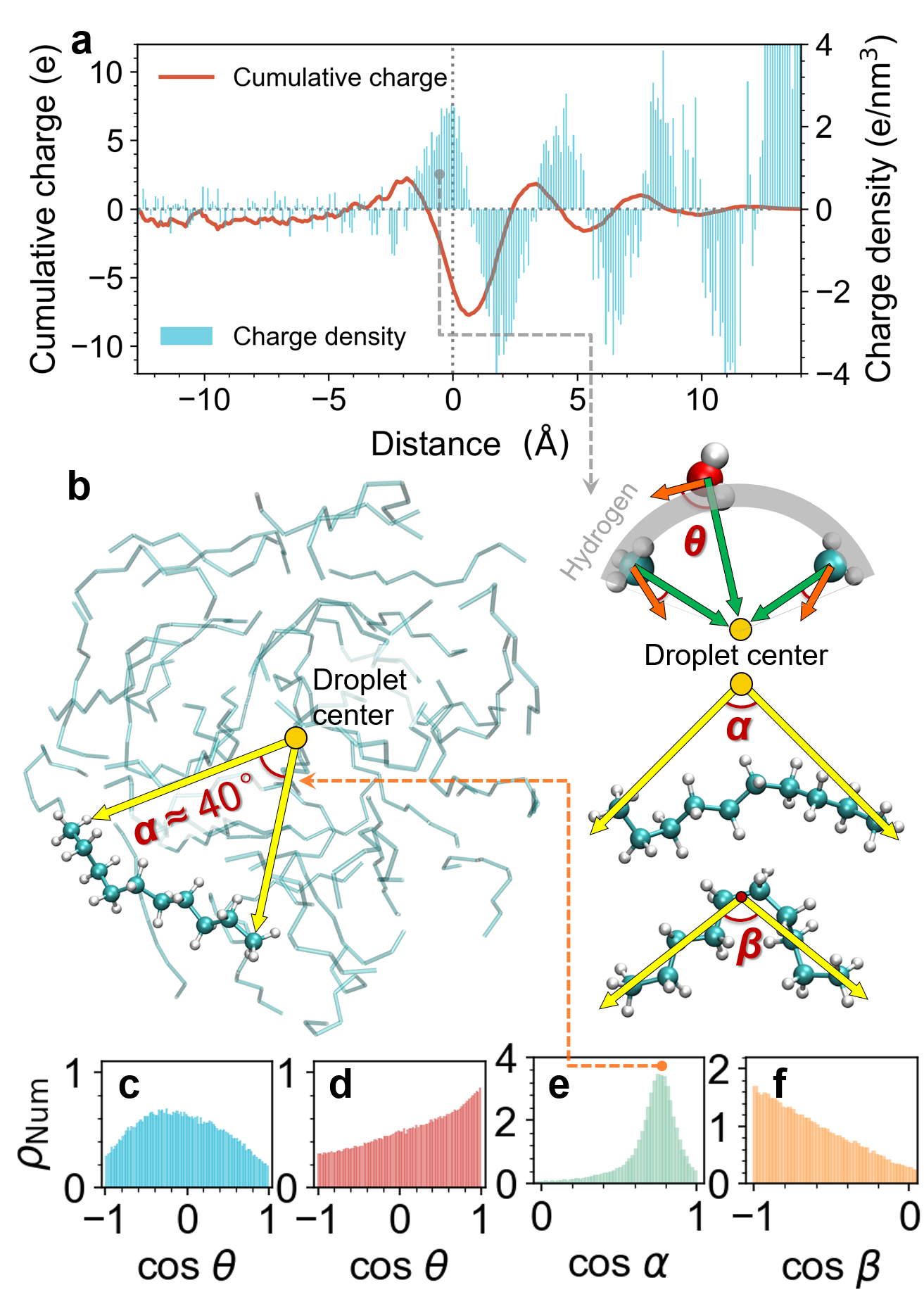}
  \caption{\textbf{Charge distribution of the oil droplet in pure water and the orientation distribution of interfacial water and decane.} \textbf{a}. Charge density and cumulative charge as functions of normal distance. The zero of normal distance is the position of interface. \textbf{b.} The highlight of a representative decade molecule that remains flat on the oil surface with an angle of \(\alpha \approx 40^\circ\), and the schematic diagrams of the angles  \(\theta\),  \(\alpha\),  and \(\beta\). \textbf{c-e.} Orientation distributions of interfacial \ce{H2O} molecules (distance greater than --0.8 Å), \ce{-CH2-} and \ce{-CH3} groups of decane (distance less than 3 Å), and decane chains. The angle \(\theta\) (as shown in \textbf{b.}) is defined as the angle between the normal vector (extending from the targeted O or C atom to the droplet center) and the dipole vector (ranging from the H center to the O or C atom). The angle \(\alpha\) (as shown in \textbf{b.}) is the angle between two vectors from the dro plet center to the two terminal carbons of decane. \textbf{f.} Bending degree of decane chains, estimated by the angle \(\beta\) (as shown in \textbf{b.}) between the two vectors from the center of the central \ce{C-C} bond to two terminal carbons of decane. }
  \label{fig_q_orien}
\end{figure*}

We next analyze the radial distribution of atomic partial charge \cite{Hirshfeld_TheorChimActa_1977_v44_p129} for the oil droplet in pure water and the preferred orientation of interfacial water and decane molecules. The Hirshfeld charge was effectively predicted by training a deep charge (DC) model \cite{Zeng_PhysRevB_2022_v105_p174109,Zhang_NpjComputMater_2024_v10_p94} with the M06-2X method. More details on the training can be found in the SI.

From the droplet center (14 Å) to the droplet surface (0 Å), a staggered distribution of positive and negative charge density is observed, as depicted in Figure \ref{fig_q_orien}a. This phenomenon probably arises from the fact that the decane molecules form the oil droplet by wrapping layer by layer around the droplet center, which is evidenced by the linearly increased \(\cos \beta\) distribution from 0 to --1 (Figure \ref{fig_q_orien}f) and the \(\cos \alpha\) distribution peaking around 0.77 (Figure \ref{fig_q_orien}e), where the former indicates that the decane molecules in the droplet are primarily straight chains and the latter shows that the outmost layer decane molecules approximately remain flat on the oil droplet surface indicated by the \(\alpha\) angle of  \(40^\circ\) as illustrated by Figure \ref{fig_q_orien}b. Consequently, along the radial direction, the hydrogen atoms with positive Hirsheld charge and the carbon atoms with negative Hirshfeld charge appear alternatively, leading to the staggered distribution of total charge density.
 
At the oil-water interface, which is defined as the position with the smallest density of system and marked by a vertical gray dot line in Figure \ref{fig_q_orien}a, there is a peak of positive charge density. Based on the orientation analysis in Figures \ref{fig_q_orien}c, d for the interfacial water molecules and the \ce{-CH2-} and \ce{-CH3} groups of interfacial decane, we infer that hydrogens of the dangling OH bonds of interfacial water layer and most hydrogens of the outmost layer decane molecules collaboratively contribute to this interfacial positive charge density peak (Figures \ref{fig_q_orien}b). The positive charge distribution at the oil-water interface may attract the \ce{OH-} ions and repel \ce{H3O+} ions. In the bulk water region ranging from --13 to --2 Å, the system approaches a nearly neutral charge density. (Figure \ref{fig_q_orien}a).

By summing the charge from the droplet center toward the water phase, we determine the cumulative charge for statistical spheres with different radii and plot them in Figure \ref{fig_q_orien}a using the red solid line. Near the oil-water interface, the cumulative charge reaches a maximum negative value of --2.6 e/nm$^3$ at the normal distance of 0.6 Å, which may explain the experimentally observed negative zeta potentials of oil droplets.\cite{creux2009strong, marinova1996charging, beattie2004pristine, noam2016protons}  Additionally, the attraction of \ce{OH-} ions by the interfacial positive charge density may also contribute to the observed negative zeta potentials.

\section*{3 Conclusion}

In summary, with enhanced sampling techniques that can efficiently sample water autoionization processes and describe well ionic diffusion and recombination, we investigated the interfacial distribution propensity of \ce{OH-} and \ce{H3O+} in air(oil)-water interfaces by deep potential molecular dynamics simulations. We first validated the double-layer distribution of water self-ions in the air-water interface with \ce{H3O+} enriching in the topmost water layer and \ce{OH-} staying in a few-angstroms deeper layer. Interestingly, we then found that different from the air-water interface, the oil(decane)-water interface attracts \ce{OH-} but excludes \ce{H3O+}. To understand the distinct behaviors, we performed the charge analysis and molecular orientation statistics for an oil droplet-water system and discovered that the layer-by-layer arrangement of nearly straight decane molecules around the droplet center allowed oil droplets in water to have staggered positive and negative charge distributions along the radial direction. The resulting positive charge distribution of the oil-water interface induces the interfacial enrichment of \(\ce{OH-}\) ions and the interface exclusion of \(\ce{H3O+}\) ions. We also noticed the negative cumulative charge of the oil droplet when involving its nearest neighboring water layer, and we speculated that this negative charge, together with the adsorbed \(\ce{OH-}\) ions, led to the experimental detection of negative zeta potentials of oil droplets. The present results enrich the knowledge on various water interfaces and offer valuable insights that contribute potentially to their application across multiple fields.
 
\section*{4 Method}

\subsection*{4.1 Neural network model training}
We aim to train three models: a deep potential (DP) model\cite{wang2018deepmdkit,zeng2023deepmd} for molecular dynamics simulations investigating water autoionization and distribution of self-ions along interfaces; a deep Kohn-Sham (DeePKS) model\cite{chen2020deepks,li2022deepksabacus} with accuracy comparable to the M06-2X functional, and the DeePKS model will be used for labeling the training data sets of energies and atomic forces needed for training the DP model; and a deep charge (DC) model\cite{Zeng_PhysRevB_2022_v105_p174109,Zhang_NpjComputMater_2024_v10_p94} for predicting atomic partial charges. To ensure diverse and uncorrelated data, we generate multiple independent MD simulations with enhanced sampling on some configuration subsystems (Table S1 in the SI).

The training sets for the three models differ significantly in size due to variations in their descriptors. All models, however, adhere to the criteria of physical symmetry and locality. The DP and DC models incorporate angular and radial atomic environment descriptors, while the DeePKS model includes gauge invariance symmetry and density matrices projected on atomic orbitals.\cite{zhang2018deeppotential,chen2020deepks,li2022deepksabacus} Consequently, the DeePKS model requires only a few hundred configurations for training, whereas the DP and DC models necessitate tens of thousands to accurately capture reactive processes as shown in Table S2 in the SI.

\subsection*{4.2 Density functional theory (DFT) calculation}
The label preparation of data sets for three models involves M06-2X\cite{zhao2008m06} calculations for energies and forces using the CP2K package\cite{kuhne2020cp2k}. These calculations utilize Goedecker-Teter-Hutter pseudopotentials\cite{goedecker1996separable,hartwigsen1998relativistic} and a QZV3P basis set\cite{vandevondele2005quickstep}. The multi-grid level employs a plane-wave cutoff of 1000 Ry for the total density and 70 Ry for the Kohn-Sham orbitals, as confirmed by a previous convergent test.\cite{zhang2023intramolecular} To speed up the convergence of self-consistent field (SCF) iterations, the auxiliary density matrix method is employed.\cite{guidon2010auxiliary}
In addition,  the Hirshfeld charge\cite{Hirshfeld_TheorChimActa_1977_v44_p129} is also calculated to prepare the data set for training the DC model.

\subsection*{4.3 Molecular dynamics simulations}
DPMD simulations are conducted using the LAMMPS software.\cite{plimpton1995fast} Prior to simulating the interface systems, the accuracy of water autoionization is assessed through 10 ns $ NVT$ (constant volume constant temperature) simulations with a 1.0 fs integration time step. 
For the air-water and oil-water slab systems, 10 ns simulations are conducted to achieve converged results. In the case of the larger oil droplet systems, 3 ns simulations are carried out. Temperature control at 300 K is achieved using a velocity-rescaling thermostat\cite{bussi2007canonical} with a damping time of 0.04 ps in all DPMD simulations. Simulation systems are listed in Table S3 in the SI.

\subsection*{4.4 Enhanced sampling settings}
In DPMD simulations for studying the autoionization and the ionic distribution, the OPES method\cite{invernizzi2020rethinking,invernizzi2022exploration,invernizzi2020unified} with Voronoi CVs is employed.\cite{grifoni2019microscopic,grifoni2020tautomeric,zhang2023intramolecular} The CVs, $\mathbf{s}$, effectively capture slow modes and infrequent occurrences by relating them to atomic coordinates. The external bias potential $V(\mathbf{s})$ is incorporated into the system's potential energy during MD simulations using the PLUMED plugin.\cite{tribello2014plumed} In OPES, $V(\mathbf{s})$ at the $n$th step is characterized by 
\begin{equation}
    V_n(\mathbf{s}) = (1-{\frac{1}{\gamma}})\frac{1}{\beta} \log (\frac{{P}_{n}{(\mathbf{s})}}{Z_n}+\epsilon),
\label{eq:V(s)}
\end{equation}
which involves the bias factor $\gamma = \beta \Delta E_{\mathrm{bias}}$, inverse Boltzmann factor $\beta = 1/{k_B T}$, the unbiased marginal distribution probability $P(\mathbf{s})$, and a normalization factor $Z$. The regularization term $\epsilon= e^{-\gamma/(1-1/\gamma)}$ ensures a positive argument for the logarithm and imposes a bias constraint that confines sampling to the specified region of interest. A value of $\Delta E_{\mathrm{bias}}$ of 75 kJ/mol is chosen to approximate the free energy barrier of water autoionization.

Based on the DPMD simulation results, the free energy surface along the CV can be calculated as follows
\begin{equation}
F(\mathbf{s}) = -\frac{1}{\beta} \log P(\mathbf{s}).
\label{eq:F(s)}
\end{equation}
In the regime of quasi-static bias, $P(\mathbf{s})$ can be reweighted as an average over the biased ensemble. \cite{invernizzi2020rethinking}

Comprehensive computational methods and additional details are available in the SI.

\section*{Data availability}
The simulation and visualization packages used to perform the calculations are freely available. The input files needed to reproduce the research are available at \url{https://github.com/Zhang-pchao/OilWaterInterface}. 

\section*{Supporting Information Available}

Potential model generation, including construction of training data sets, DFT calculation, and model training for DeePKS model, DP model, and DC models.
Molecular dynamics simulations, including those for the data set generation and for studying the autoionization and the ionic distribution.
Enhanced sampling settings, including bias potential, definition of collective variables, and free energy calculation.
Simulation workflow, schematic diagram of simulation systems, diffusion coefficients, error distribution on test data sets, free energy, orientation distribution, hydrogen bond, and other details (PDF).


\section*{Conflicts of interest}
The authors declare no competing financial interest.

\section*{Acknowledgments}
This work was partially supported by the National Natural Science Foundation of China under Grant Nos. [21973053, 11988102]. The authors acknowledge the valuable discussions with Professor Michele Parrinello, Axel Tosello Gardini, and Ling Liu. Computational resources were provided by the High Performance Computing (HPC) platform at Tsinghua University, the HPC Franklin at Fondazione Istituto Italiano di Tecnologia, and the Bohrium Cloud Platform (\url{https://bohrium.dp.tech}) from DP Technology.

\renewcommand{\bibfont}{\footnotesize\linespread{0.9}\selectfont}
\bibliography{references}
\end{multicols}
\newpage

\section*{\centering Table of contents (TOC)}

\begin{figure}
  \centering
  \includegraphics[width=0.6\textwidth]{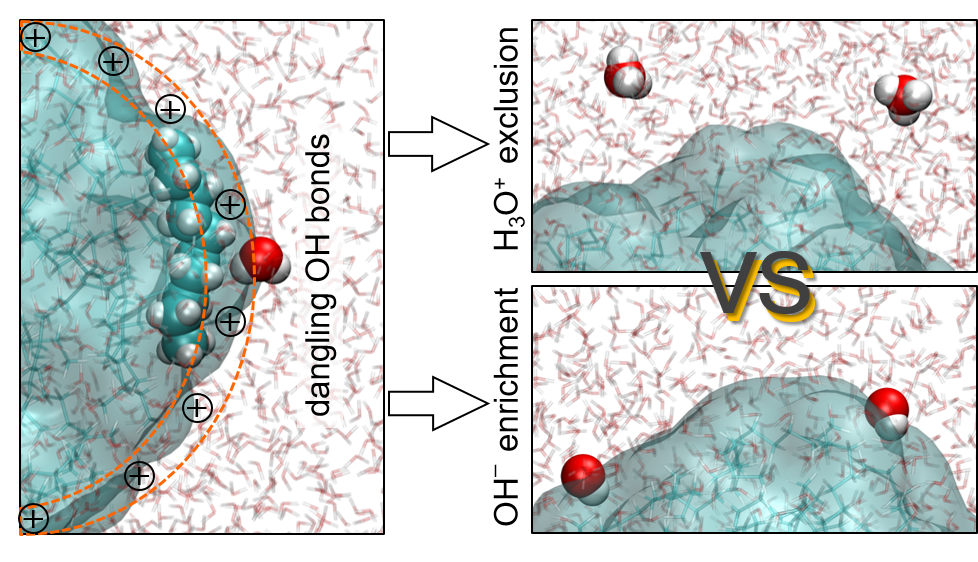}
  \label{fig_toc}
\end{figure}

\end{document}